\documentclass[final]{svjour3}
\usepackage{graphicx} 
\usepackage{rotating}
\usepackage{amssymb}
\usepackage{mathptmx}
\usepackage[numbers]{natbib}
\makeatletter
\journalname{Journal of Low Temperature Physics}

\usepackage{subfigure}

\begin{document}

\newcommand{\hdblarrow}{H\makebox[0.9ex][l]{$\downdownarrows$}-}

\title{Hierarchical sinuous phased array for increased mapping speed of multichroic focal planes}

\author{Ari Cukierman$^1$ \and Adrian T. Lee$^{1,2,3}$ \and Christopher Raum$^2$ \and Aritoki Suzuki$^{1,2,3}$ \and Benjamin Westbrook$^2$}

\institute{
$^1$ Department of Physics, University of California, Berkeley \\
$^2$ Radio Astronomy Laboratory, University of California, Berkeley \\
$^3$ Physics Division, Lawrence Berkeley National Laboratory \\
\email{ajcukierman@berkeley.edu}
}

\maketitle

\begin{abstract}

We present the design, fabrication and measured performance of a hierarchical sinuous-antenna phased array coupled to transition-edge-sensor (TES) bolometers for measurements of the cosmic microwave background (CMB). To efficiently cover a broader range of frequencies, many CMB experiments have begun to deploy multichroic pixels. For a given cryogenic and optical design, the pixel size that optimizes the mapping speed is frequency-dependent and increases with wavelength. If a multichroic pixel is the same size at all frequencies, then the mapping speed will be optimal in at most one frequency band. To achieve optimal mapping speed at all frequencies, a pixel size that varies with frequency is needed. This can be accomplished by creating phased arrays from neighboring pixels, where the size of each phased array is chosen independently for each frequency band. As the array unit, we choose a lenslet-coupled sinuous antenna on account of its wide bandwidth, dual polarization and constant beam waist. We find that the array factor tends to compensate for beam non-idealities, so that the sinuous antenna can be substantially undersized relative to the single-element case; this frees up valuable focal-plane area for bolometers, bandpass filters, summing networks, readout circuitry, etc. An additional benefit of a hierarchical phased array is the reduced readout requirement: since the low frequencies are arrayed to increase their mapping speeds, the detector count scales approximately logarithmically with the number of frequency bands. Here we present measurements from a prototype device, in which hierarchical triangular arrays are used at 90, 150 and 220 GHz to keep the effective pixel size and, therefore, the beam width roughly constant across the entire frequency range. The fabrication process is described, and the utility of hierarchical phased arrays is discussed in the context of upcoming CMB experiments.

\keywords{cosmic microwave background, sinuous antenna, transition edge sensor, phased array, hierarchical, multiscale, multichroic, bolometer, lenslet}

\end{abstract}

\section{Increasing mapping speed}

\begin{figure}[h]
\begin{center}
\subfigure{\includegraphics[height = 0.25\textwidth]{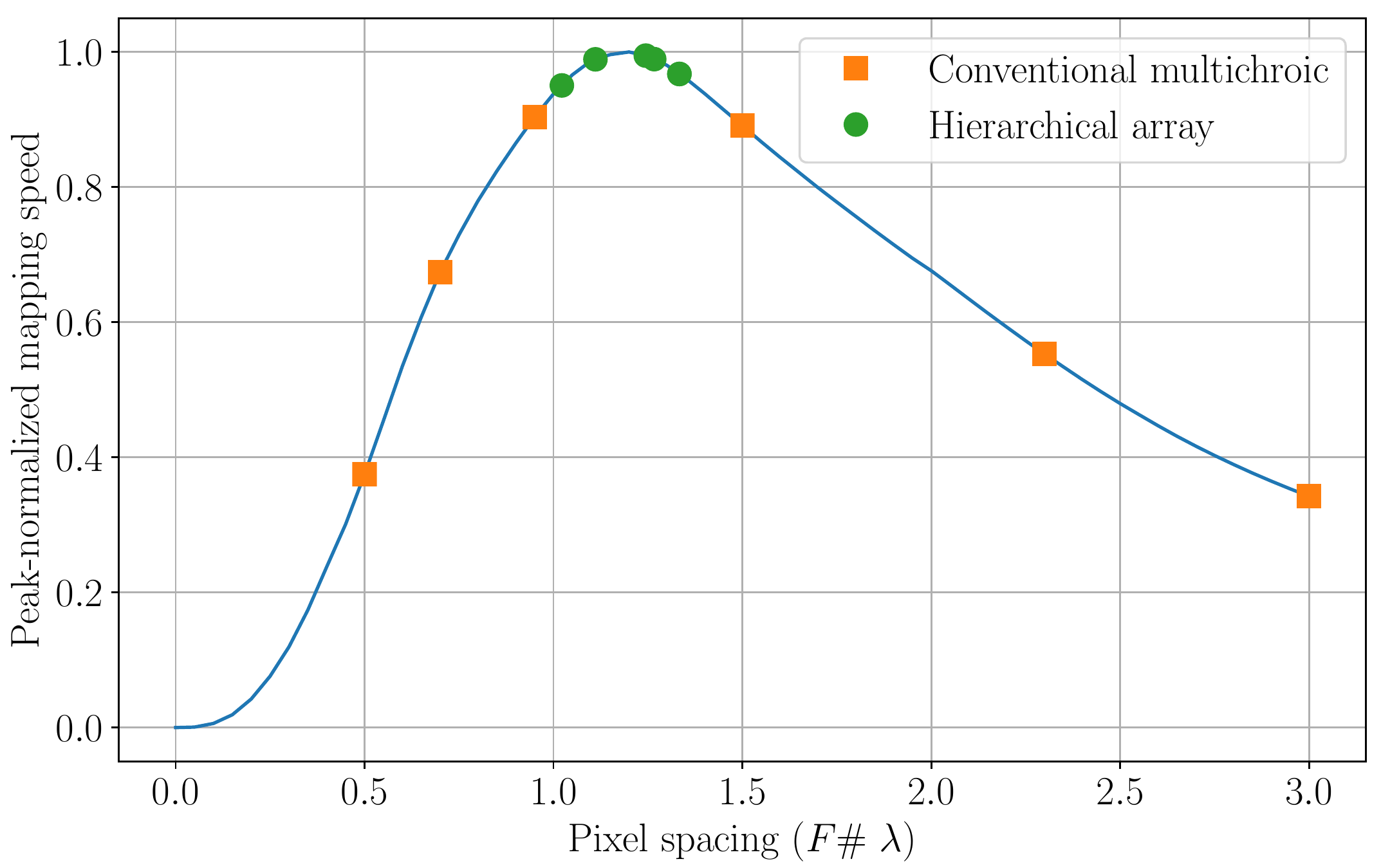}}
\hspace{0.01\textwidth}
\subfigure{\includegraphics[height = 0.25\textwidth]{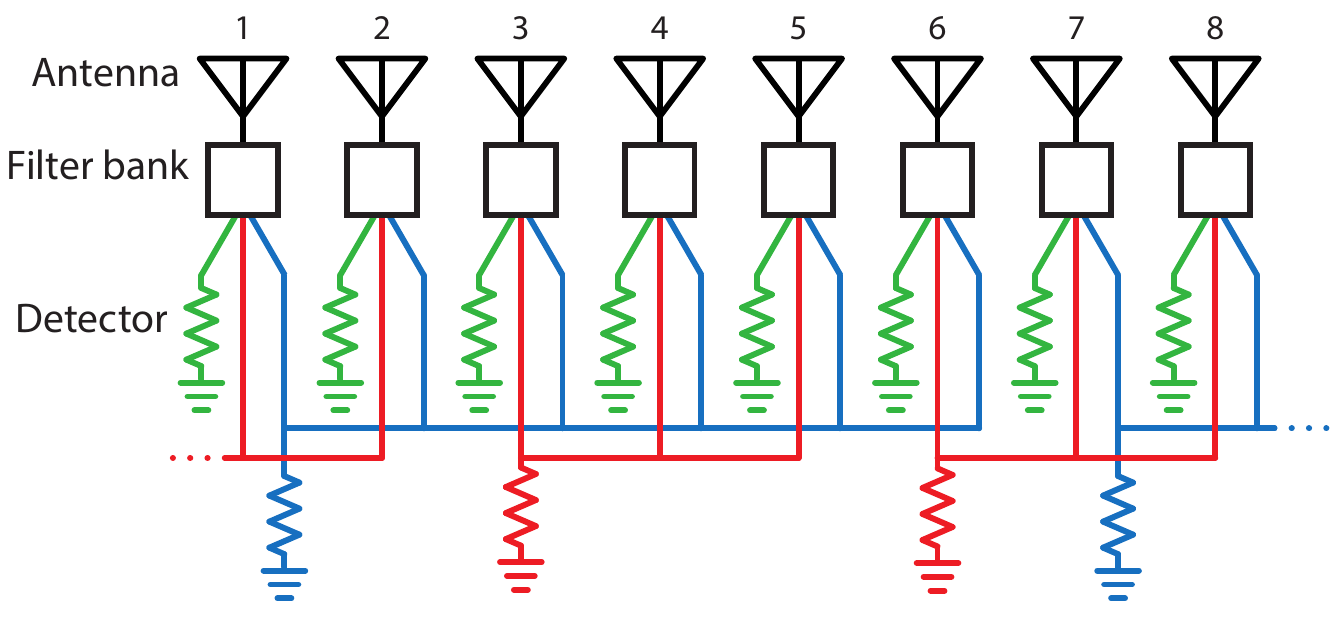}}
\end{center}
\caption{\emph{Left} Mapping speed as a function of pixel spacing (\emph{blue solid line}) in units of f-number of the detector-aperture system times wavelength ($F\#~ \lambda$) for an example multichroic focal plane. The \emph{square orange markers} indicate where a conventional multichroic pixel would fall on this curve in each of its 6 frequency bands. The \emph{green circular markers} indicate the same for a hierarchical focal plane; they are much closer to the optimum in all frequency bands. \emph{Right} Topology of an example hierarchical focal plane. The highest frequencies (\emph{green}) dump power to bolometers in a conventional way. The middle frequencies (\emph{red}) sum 3 antennas coherently before dumping power to a bolometer. The lowest frequencies (\emph{blue}) sum 6 antennas coherently before dumping power to a bolometer. }
\label{fig:mappingSpeed and topology}
\end{figure}

For a given experimental configuration, the mapping speed depends on the pixel spacing. Due to the competing concerns of detector number and aperture loading, the mapping speed typically has a clear maximum. The optimal pixel spacing, however, is approximately proportional to the wavelength. For a focal plane composed of multichroic pixels, then, the spacing is optimal at at most one frequency. The blue points in Fig.~\ref{fig:mappingSpeed and topology} (\emph{left}) show the mapping speeds for the bands of a hexachroic pixel. Several bands fall substantially below the optimal mapping speed.

This problem can be solved with a pixel size that varies with frequency. If this were realized, then the optimal pixel spacing could be chosen independently for each frequency band. The red points in Fig.~\ref{fig:mappingSpeed and topology} (\emph{left}) show the mapping speeds for the bands of such a \emph{multiscale} or \emph{hierarchical} focal plane. In this case, all of the frequency bands are close to the optimal mapping speed.

How can a multiscale or hierarchical multichroic focal plane be realized? We propose a hierarchical phased array composed of ultrawideband antenna elements. Each antenna element receives a broad range of frequencies which are then split into frequency bands via multiplexing bandpass filters. At this point, the transmission lines carrying different frequencies can be treated independently (except for possible crossovers). We coherently sum the signals from neighboring antennas to synthesize antenna arrays, which act as effectively larger pixels. The size of each array is chosen differently for each frequency band. Figure~\ref{fig:mappingSpeed and topology} (\emph{right}) shows the topology of a 3-band hierarchical phased array.

\begin{figure}[h]
\begin{center}
\subfigure{\includegraphics[height = 0.25\textwidth]{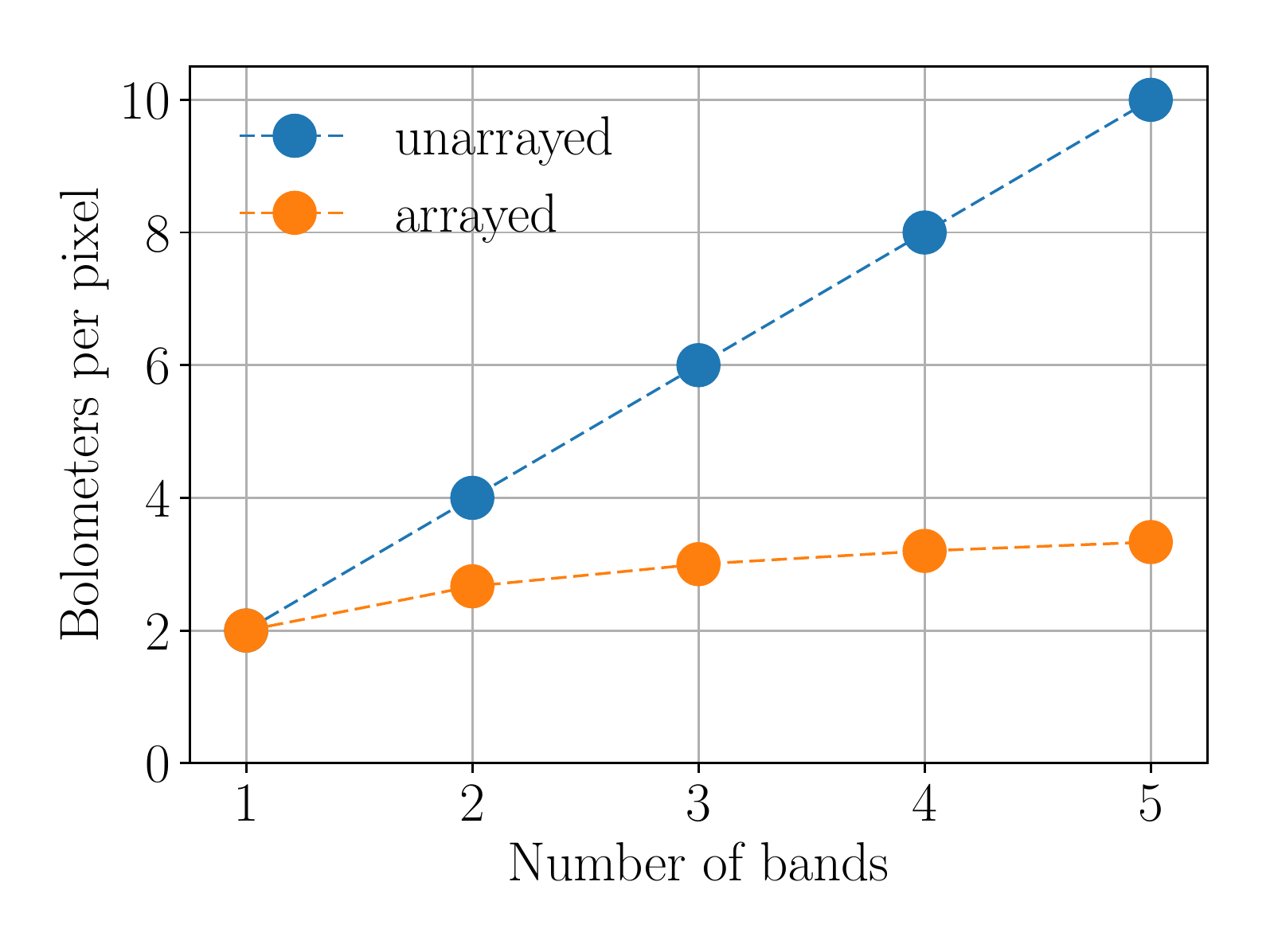}}
\hspace{0.01\textwidth}
\subfigure{\includegraphics[height = 0.25\textwidth]{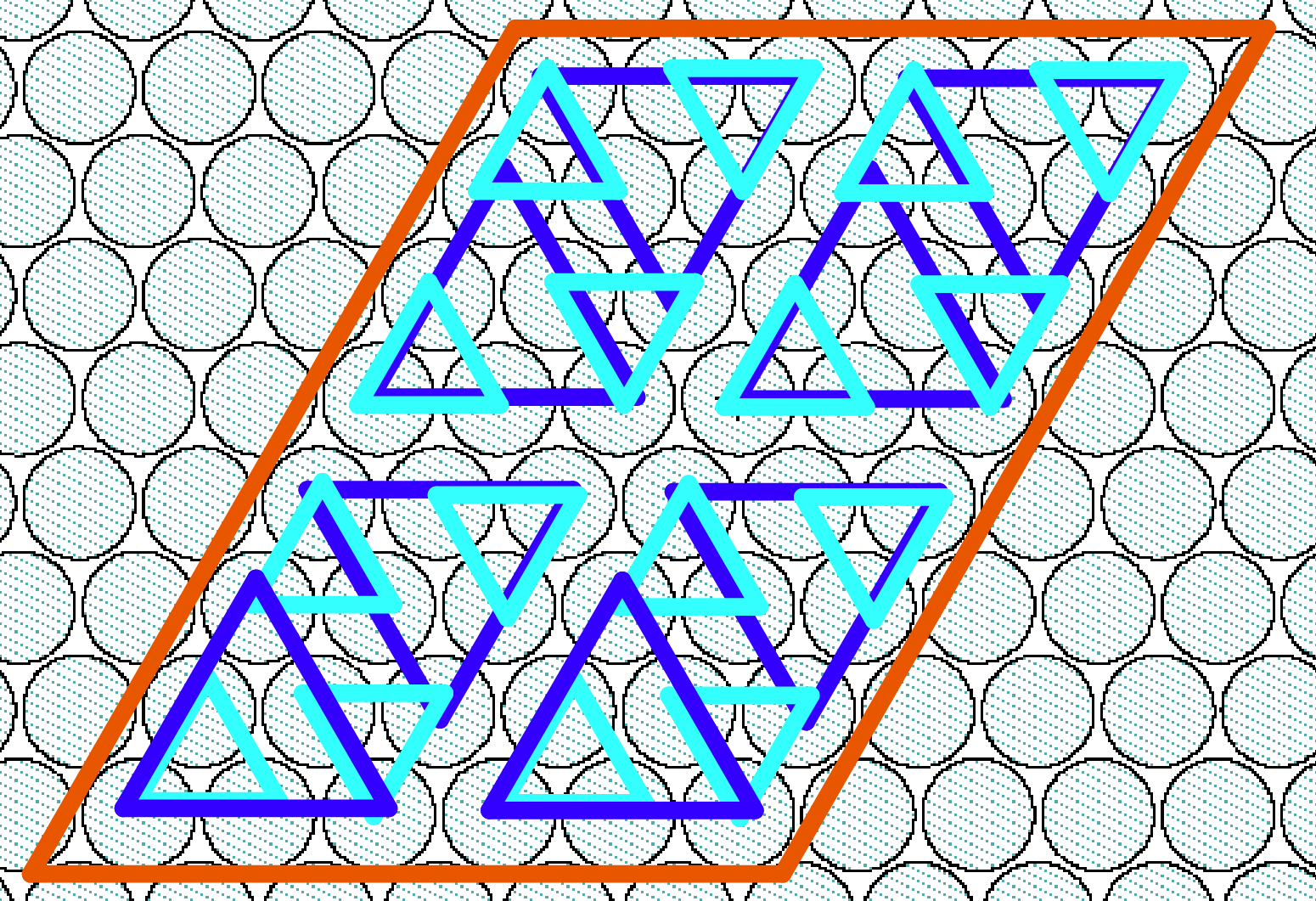}}
\end{center}
\caption{\emph{Left} Comparison of the number of readout channels for a conventional multichroic focal plane (\emph{blue}) and a hierarchical focal plane (\emph{orange}). The required readout increases linearly with the number of bands in the former case but only logarithmically in the latter case. \emph{Right} Example wafer tiling for a 3-level hierarchy on a hexagonal close-packed grid. The triangles form rhombi, which tile the plane. }
\label{fig:readoutSavings and rhombus}
\end{figure}

A fortunate consequence of a multiscale architecture is that the number of detectors actually \emph{decreases}, although the mapping speed \emph{increases}.  This is because the effective pixel spacing is larger at lower frequencies: for a fixed focal-plane area, there are, therefore, fewer effective pixels.
For a non-hierarchical multichroic focal plane, the number of detectors increases linearly with the number of frequency bands. In the hierarchical case, the increase is approximately logarithmic.
Figure~\ref{fig:readoutSavings and rhombus} (\emph{left}) shows the decrease in the number of bolometers for a triangularly hierarchical focal plane. Already at 3 bands, the decrease in the number of readout channels is a factor of 2; at 5 bands, it is a factor of 3.

\section{Prototype}

A prototype was built using lenslet-coupled sinuous antennas \cite{Westbrook:2016bkt, Edwards:2012} as the basic elements of the hierarchy. The advantages of this antenna are its wide bandwidth and constant beam waist. This latter feature ensures that grating lobes are minimized even in the higher-frequency phased arrays. The detectors are transition-edge-sensor (TES) bolometers.

A 3-level triangular hierarchy was designed using individual antennas at 220 GHz, 3-pixel triangles at 150 GHz and 6-pixel triangles at 90 GHz. This prototype was designed for a 20x20 mm die and, therefore, did not include a complete hierarchical tiling of the wafer. The prototype wafer was fabricated at the Marvell Nanolab at UC Berkeley using a process similar to that of the POLARBEAR-2 detector wafers \cite{Westbrook:2016bkt}.

\begin{figure}[h]
\begin{center}
\subfigure{\includegraphics[height = 0.25\textwidth]{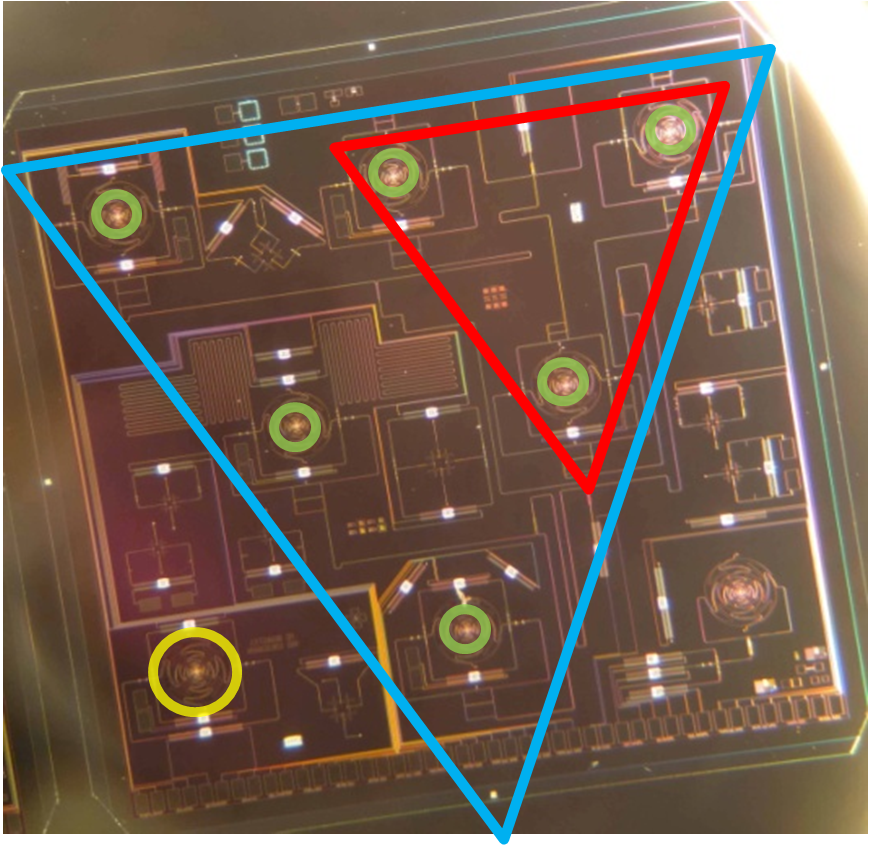}}
\hspace{0.01\textwidth}
\subfigure{\includegraphics[height = 0.25\textwidth]{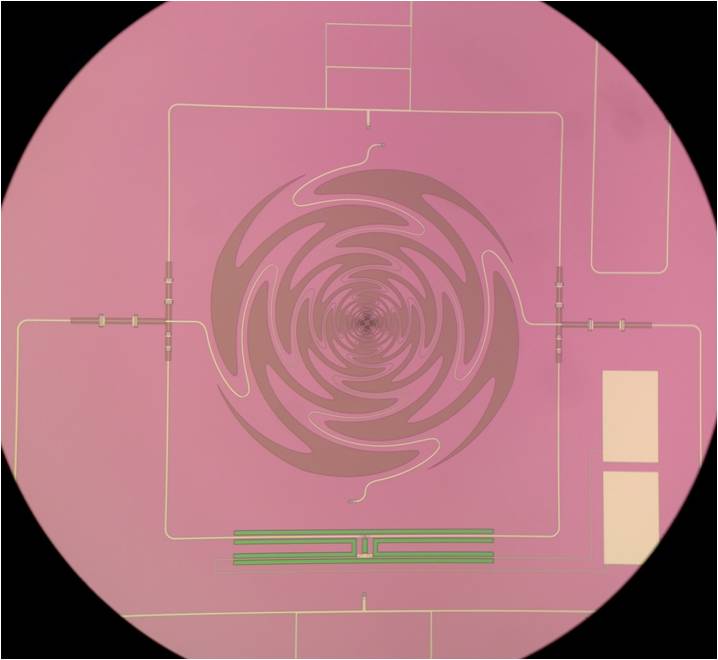}}
\hspace{0.01\textwidth}
\subfigure{\includegraphics[height = 0.25\textwidth]{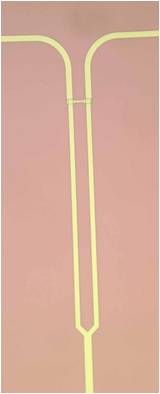}}
\hspace{0.01\textwidth}
\subfigure{\includegraphics[height = 0.25\textwidth]{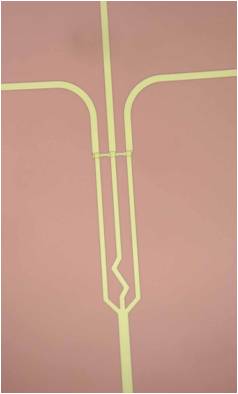}}
\hspace{0.01\textwidth}
\subfigure{\includegraphics[height = 0.15\textwidth]{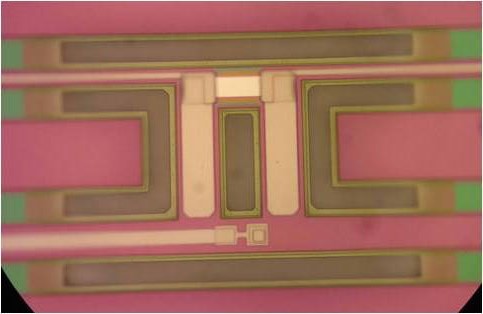}}
\hspace{0.01\textwidth}
\subfigure{\includegraphics[height = 0.15\textwidth]{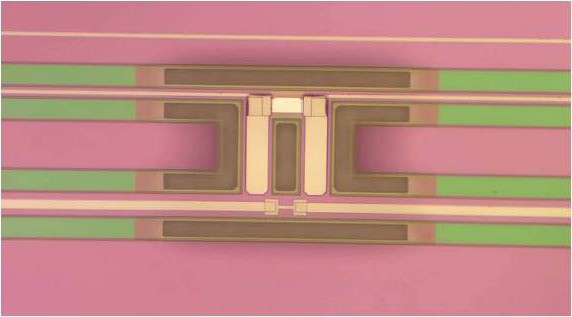}}
\end{center}
\caption{\emph{Upper, leftmost} Wafer die on which a hierarchical phased array is defined. The \emph{yellow} circle in the bottom left indicates a conventional trichroic pixel. The small \emph{green} circles indicate pixels for which the 220-GHz signal is dumped directly to a bolometer without coherent summing. The \emph{red} triangle covering 3 pixels indicates the 150-GHz array, i.e., the 150-GHz signals have been coherently summed before dumping power to a bolometer. The \emph{blue} triangle covering 6 pixels indicates the 90-GHz array, i.e., the 90-GHz signals have been coherently summed before dumping power to a bolometer. \emph{Upper, second from left} Unit cell of the hierarchical array. This pixel is sensitive to a single polarization for simplicity of design and fabrication. The sinuous antenna is fed with balanced microstrip lines. At the left and right of the antenna, there are trichroic bandpass filters. The balanced microstrip lines are combined at the top and bottom in double ring hybrids to form unbalanced microstrip lines, which allow for a simpler summing network. The 220-GHz signal is dumped directly to the bolometer just below the sinuous antenna. \emph{Upper, second from right} A 2-stage Wilkinson divider for coherently summing 2 microstrip signals. \emph{Upper, rightmost} A 2-stage planarized Wilkinson ``trivider\rq{}\rq{} for coherently summing 3 microstrip signals. \emph{Bottom left} Bolometer island for an unbalanced microstrip line that dumps power to a Ti resistor connected by via to the ground plane. \emph{Bottom right} Bolometer island for balanced microstrip lines that dump power to a differentially fed Ti resistor. }
\label{fig:wafer}
\end{figure}

The phased array is composed of unit cells with the following structure, which is partially depicted in Fig.~\ref{fig:wafer}. A lenslet-coupled sinuous antenna receives the incident microwave radiation. The sinuous antenna can be significantly undersized relative to the single-pixel multichroic case, since the beam shape is determined in large part by the phased-array geometry. Here we use a 1.7-mm diameter, cf. 3 mm for SPT3G (South Pole Telescope 3G) 90/150/220-GHz trichroic pixels \cite{Benson:2014qhw}. The signal from the antenna is split into 3 frequency bands centered at 90, 150 and 220 GHz. The 220-GHz signal is immediately terminated on a bolometer. The balanced 90- and 150-GHz signals are combined in $180^\circ$ hybrids to form unbalanced microstrip lines, which then enter the summing network.

The summing network is designed so that the length of each line is the same coming from each antenna element. This ensures that the phases are the same when the signals are coherently summed. Microstrip meanders are used to force these lengths to be equal. 
The signals are combined using Wilkinson dividers and ``trividers\rq{}\rq{}, where the names refer to the reverse-time performance. We use 2-stage Wilkinson splitters to increase the bandwidth. The 3-pixel triangle uses a single trivider; the 6-pixel triangle uses both a divider and a trivider.

The microstrip lines exiting the sinuous antenna are balanced, so we terminate the 220-GHz lines \emph{differentially} on a lumped resistor on the bolometer island. The 90- and 150-GHz signals are converted to a single \emph{unbalanced} microstrip line, which we terminate on a single-ended lumped resistor connected to ground with a via. For optical testing, the bolometer legs were designed to be short in order to increase the saturation power.


The bolometer and antenna wafer is aligned and coupled to a lenslet array, which is composed of antireflection-coated silicon extended hemispherical lenses. For this prototype, a POLARBEAR-1 lenslet array was used for testing \cite{Kermish:2012eh}. Although this lenslet array was designed for use at 150 GHz only, its performance at 90 and 220 GHz is good enough to test the main features of a hierarchical sinuous phased array. 

\begin{figure}[h]
\begin{center}
\includegraphics[height = 0.25\textwidth]{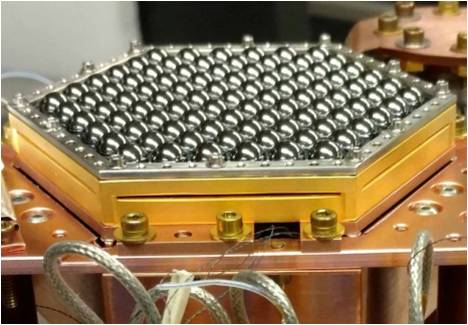}
\end{center}
\caption{The wafer module used for testing is composed mainly of components from POLARBEAR-1 \cite{Kermish:2012eh}. The prototype wafer was designed to mate with a POLARBEAR-1 lenslet array in an invar holder. The mK mounting and wiring hardware was designed to mate with this invar holder. }
\label{fig:testSetup}
\end{figure}

The device and lenslet wafers are clamped together in an invar holder, which is in turn bolted to the 250-mK stage of a He3 sorption refrigerator in an 8-inch IRLabs dewar (Fig.~\ref{fig:testSetup}). The devices were tested at UC Berkeley using a thermal source on an XY stage to measure beams and a thermal source on a Fourier-transform spectrometer (FTS) to measure spectra.

\begin{figure}[h]
\begin{center}
\subfigure{\includegraphics[height = 0.25\textwidth]{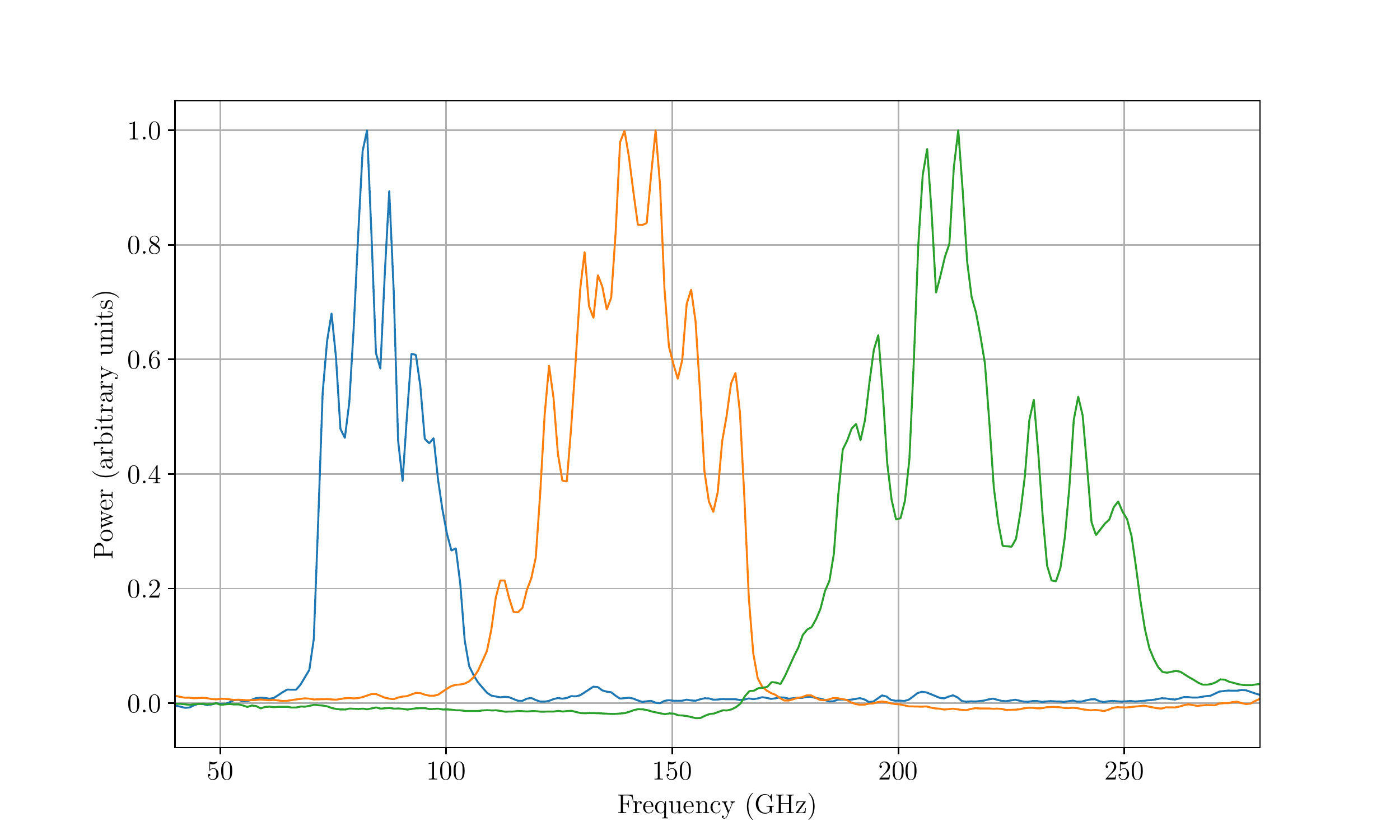}}
\hspace{0.01\textwidth}
\subfigure{\includegraphics[height = 0.25\textwidth]{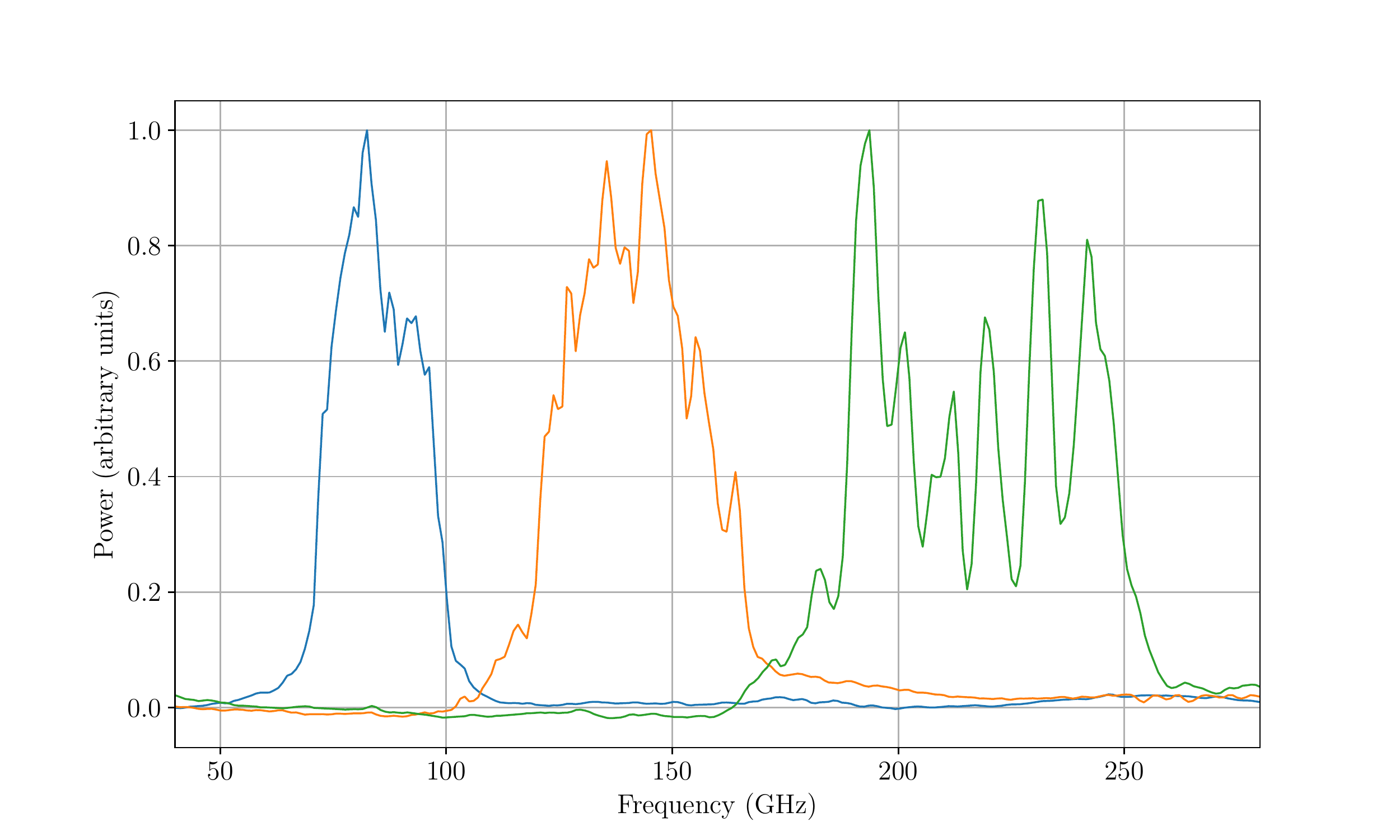}}
\end{center}
\caption{\emph{Left} Frequency response for the conventional (unarrayed) trichroic pixel. \emph{Right} Frequency response for the hierarchical array. }
\label{fig:bands}
\end{figure}

\begin{figure}[h]
\begin{center}
\subfigure{\includegraphics[height = 0.25\textwidth]{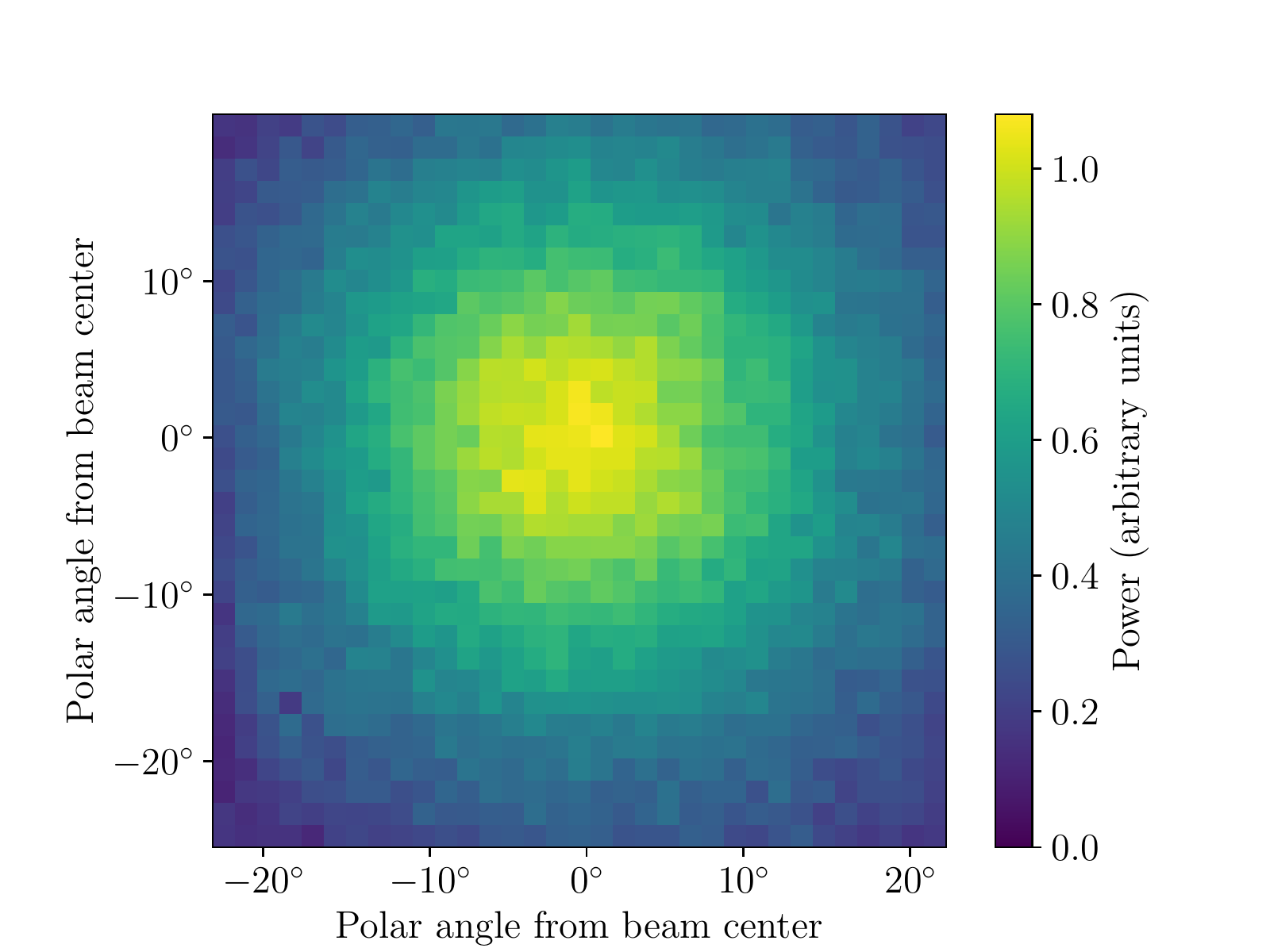}}
\hspace{0.01\textwidth}
\subfigure{\includegraphics[height = 0.25\textwidth]{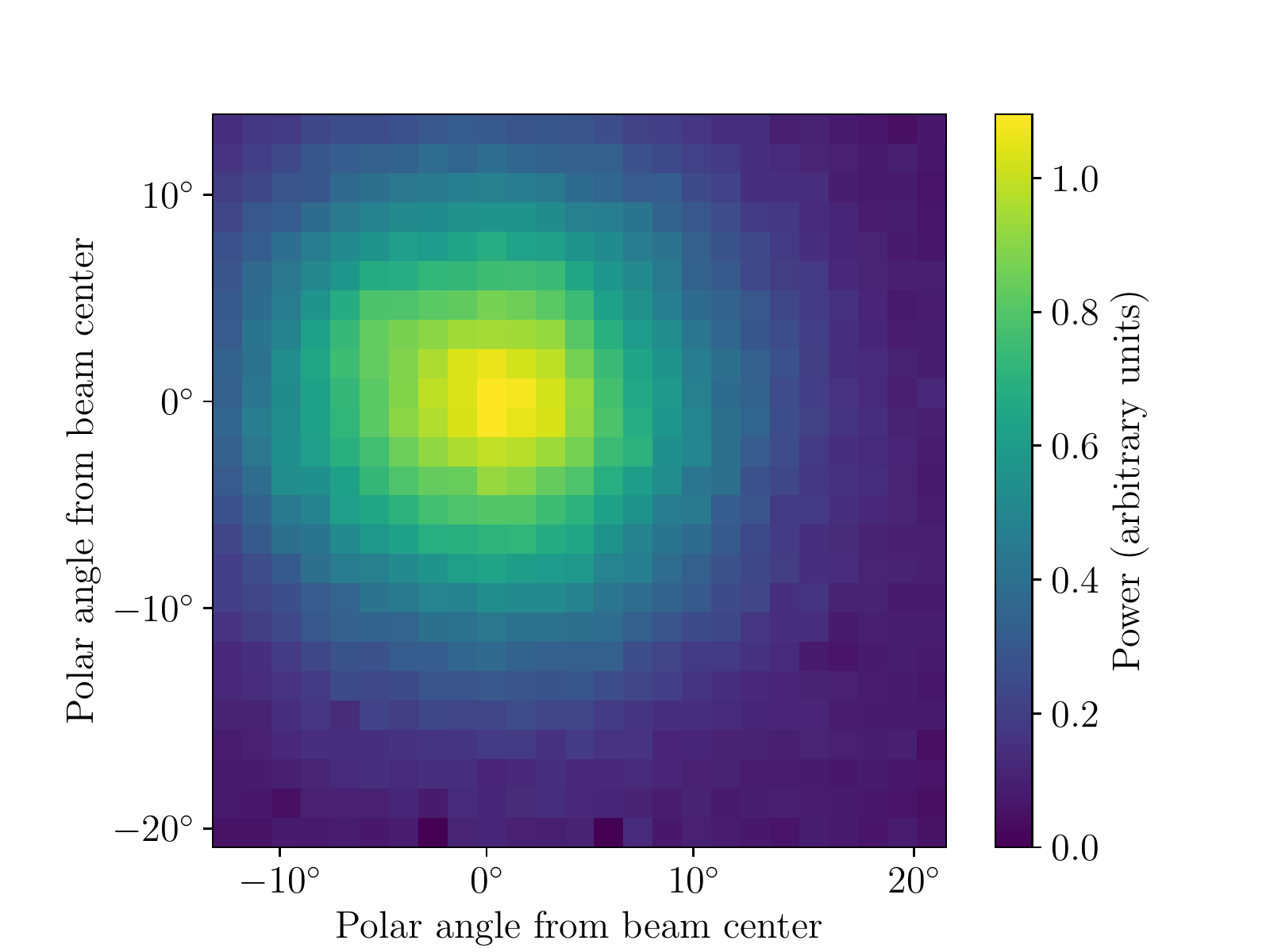}}
\hspace{0.01\textwidth}
\subfigure{\includegraphics[height = 0.25\textwidth]{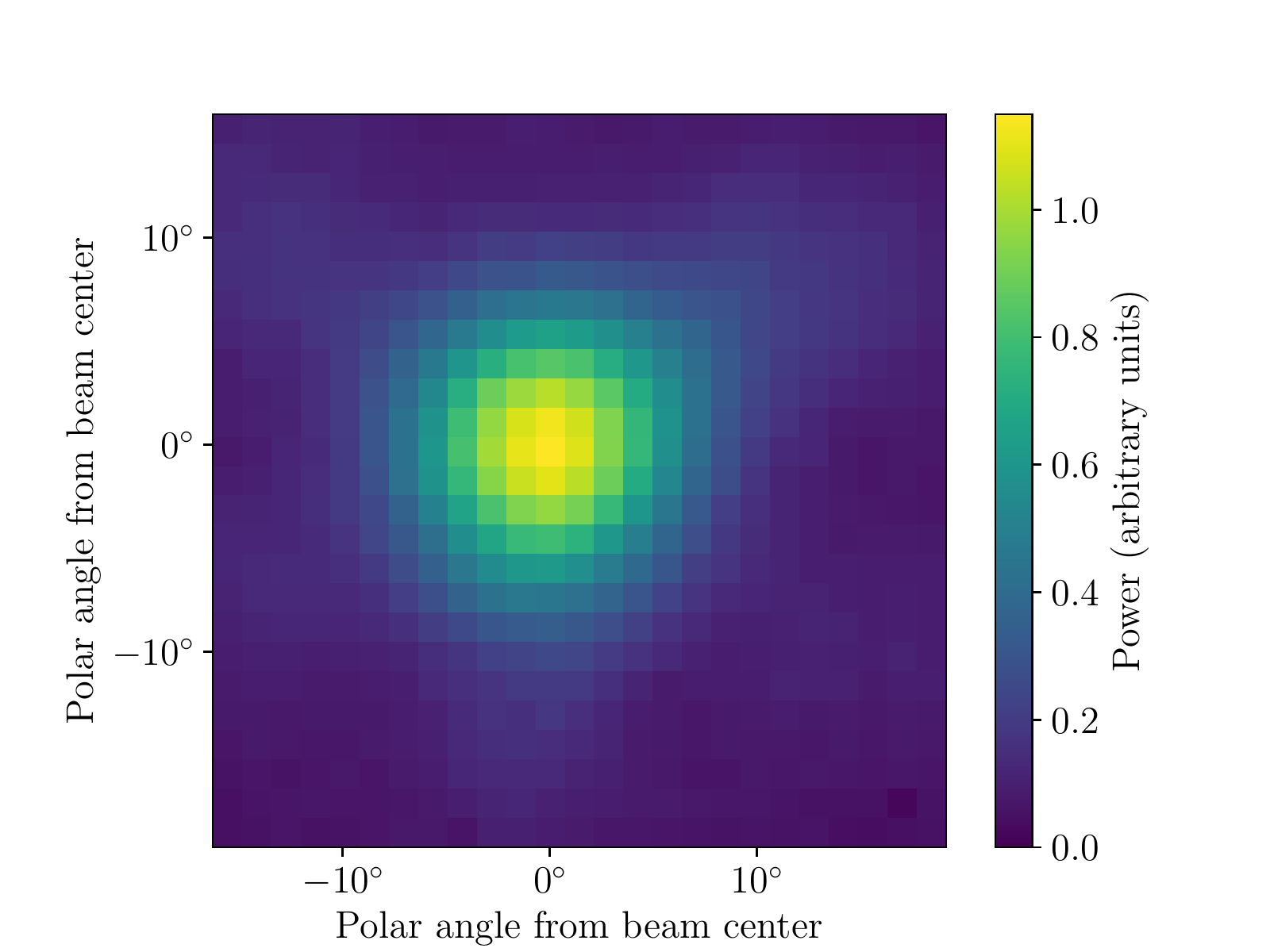}}
\hspace{0.01\textwidth}
\subfigure{\includegraphics[height = 0.25\textwidth]{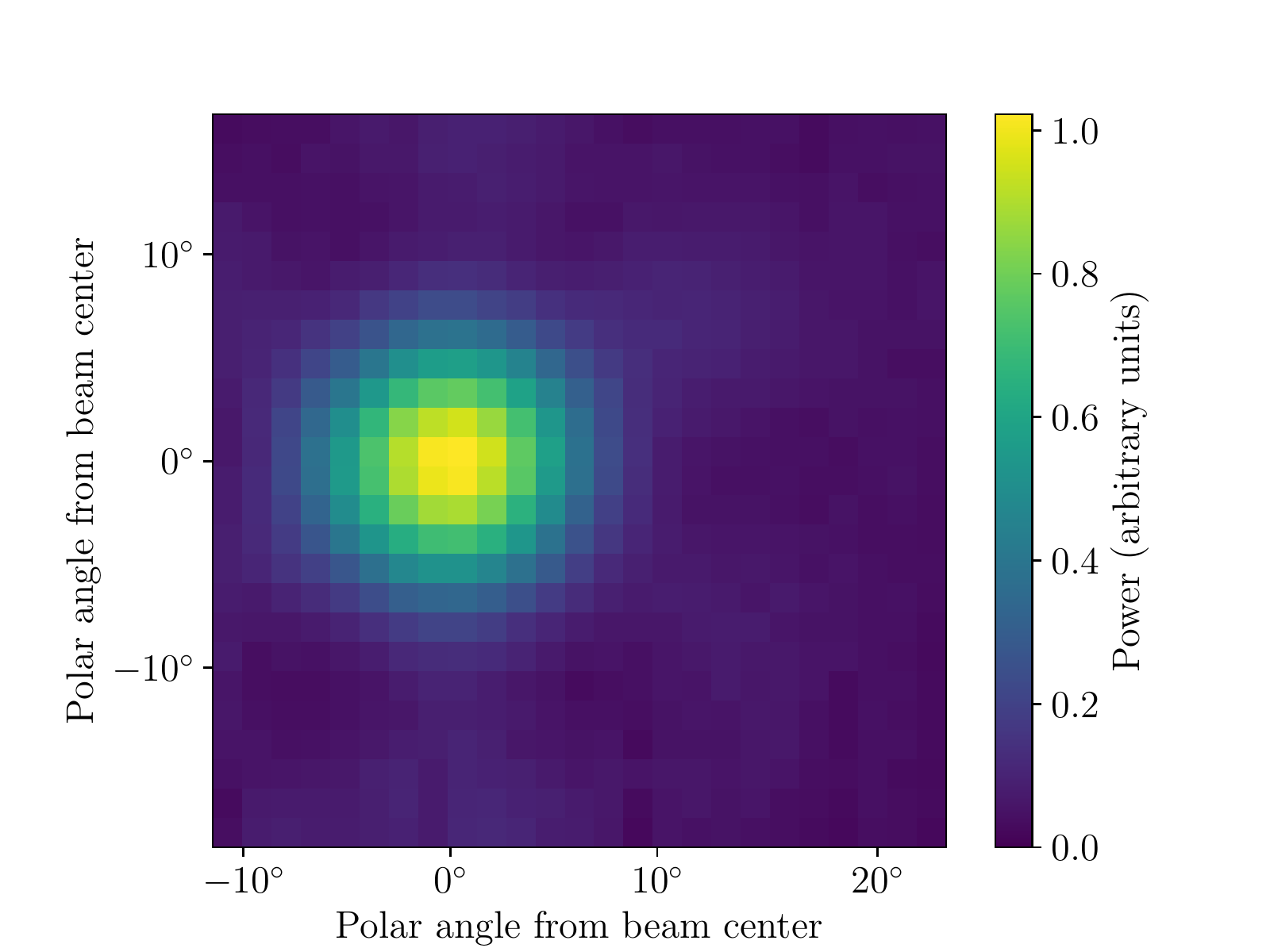}}
\end{center}
\caption{Beam maps of the 90- and 150-GHz devices. \emph{Upper left} Unarrayed (single-pixel) 90-GHz beam. \emph{Upper right} Unarrayed (single-pixel) 150-GHz beam. \emph{Lower left} Arrayed (6-pixel) 90-GHz beam. \emph{Lower right} Arrayed (3-pixel) 150-GHz beam. The offsets among these beam centers are expected based on the offsets of the phase centers on the device wafer. The faint hexagonal streaks in the arrayed beams are expected from the array factor for triangular antenna arrays. }
\label{fig:beamMaps}
\end{figure}

\begin{figure}[h]
\begin{center}
\subfigure{\includegraphics[height = 0.25\textwidth]{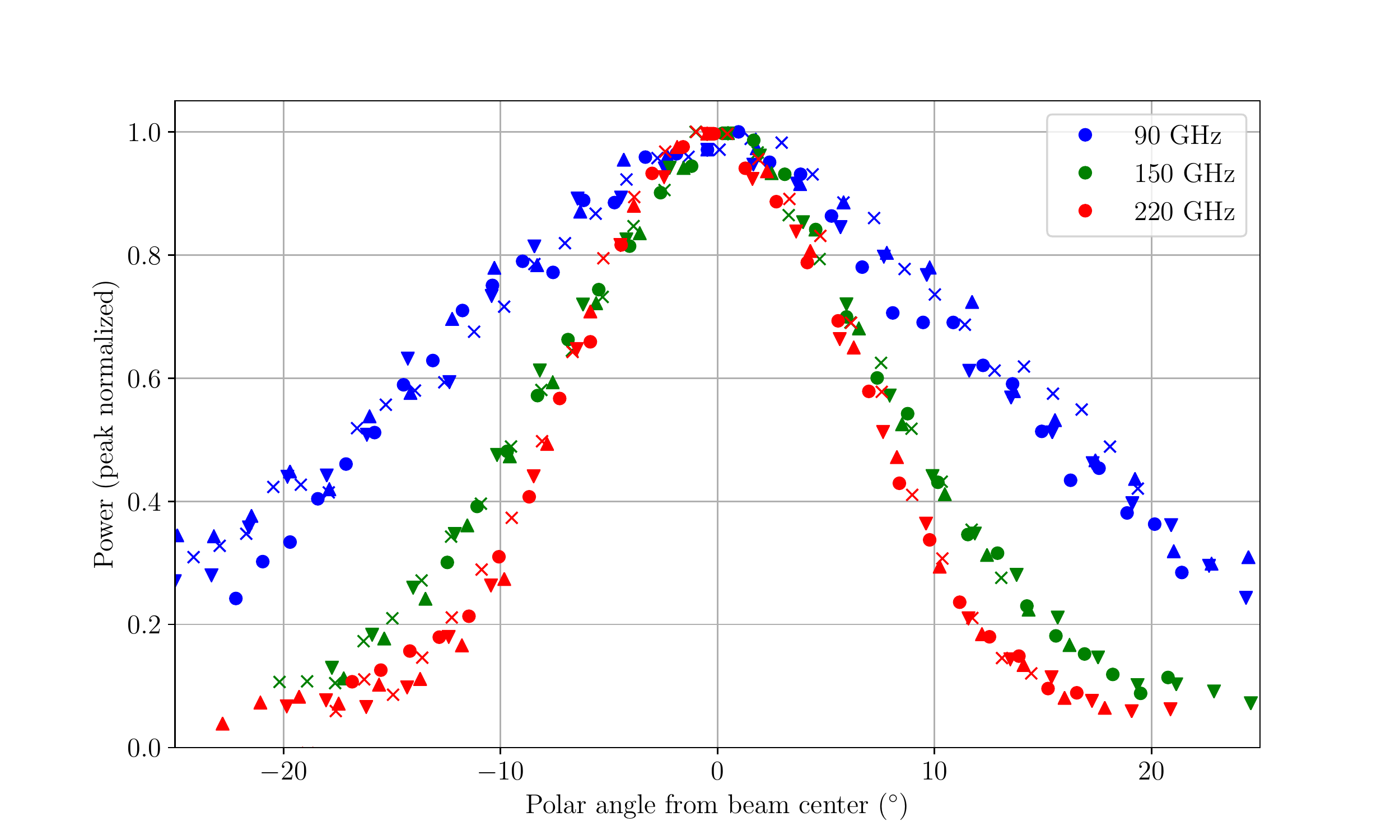}}
\hspace{0.01\textwidth}
\subfigure{\includegraphics[height = 0.25\textwidth]{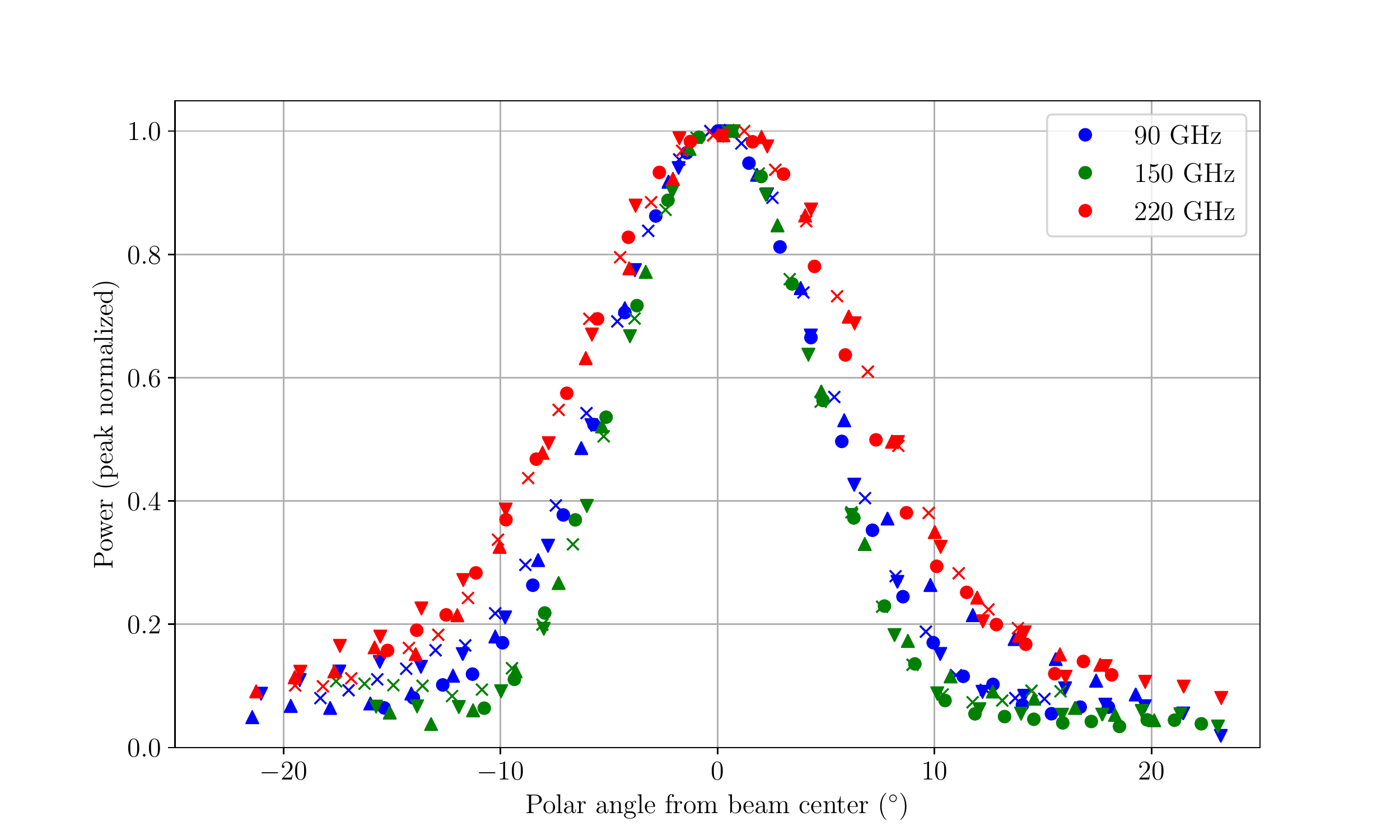}}
\end{center}
\caption{ Beam cuts for the 90-, 150- and 220-GHz devices. The colors indicate frequency bands; the marker styles indicate cut planes ($0^\circ$, $45^\circ$, $90^\circ$ and~$135^\circ$). \emph{Left} Beam cuts for an unarrayed trichroic pixel. \emph{Right} Beam cuts for hierarchically arrayed pixels. In the latter case, the beams are more similar in size and also rounder. }
\label{fig:beamCuts}
\end{figure}

As a kind of control, a single-pixel trichroic pixel was measured from the same wafer. This pixel shows 3 well-defined frequency bands (Fig.~\ref{fig:bands} \emph{left}) and beam widths that decrease inversely with frequency (Fig.~\ref{fig:beamMaps} \emph{upper} and Fig.~\ref{fig:beamCuts} \emph{left}). 
The hierarchical phased array shows 3 similar frequency bands (Fig.~\ref{fig:bands} \emph{right}) with beam widths that are approximately constant with frequency (Fig.~\ref{fig:beamMaps} \emph{lower} and Fig.~\ref{fig:beamCuts} \emph{right}). The arrayed beams display an azimuthal symmetry driven by the array factor.

Despite the short bolometer legs, the bolometers were saturated which precluded any measurement of efficiency. Non-linearity was removed from the FTS measurements. Future iterations will use silicon nitride as a microstrip dielectric as opposed to the silicon oxide used here. The demonstrated low loss of silicon nitride at mm wavelengths will lessen any concerns about the long microstrip lengths necessary for the summing network.
The steering of the beams is constrained to less than 1 degree. Faint hexagonal streaks can be seen in the beam maps; these are expected from the array factor of triangular antenna arrays.

\section{Future directions}

What would a hierarchical focal plane look like? We want to pack the antenna elements as closely as possible to receive as many photons as possible. For this reason, we choose a hexagonal close packing as the underlying tiling of the antenna elements. The hexagonal packing lends itself to both hexagonal and triangular arrays. Triangular arrays tend to have more convenient numbers of elements, e.g., 3, 6, 10, etc. as opposed to 7, 19, 37, etc. 
We would like the tiling to repeat in a similar way for all array sizes and to be contained within the same area. A possible realization of a 3-level hierarchy is shown in Fig.~\ref{fig:readoutSavings and rhombus} (\emph{right}), i.e., a rhomboidal tiling. In this case, the wafer is divided into a repeating pattern of 2 large triangles and 4 small triangles. 

Experiments that have deployed or are planning to deploy lenslet-coupled sinuous antennas include SPT3G \cite{Benson:2014qhw}, Simons Array \cite{Suzuki:2015zzg} and LiteBIRD \cite{Matsumura:2016sri}. A hierarchical scheme could be implemented as an upgrade to these experiments or in the next generation of experiments.

\begin{acknowledgements}
This work was supported by NASA grant NNX17AH13G. The devices were fabricated in the Marvell Nanolab at UC Berkeley.
\end{acknowledgements}

\pagebreak


\begin{thebibliography}{99}

\bibitem{Westbrook:2016bkt} 
  B.~Westbrook, A.~Cukierman, A.~Lee, A.~Suzuki, C.~Raum and W.~Holzapfel,
  J.\ Low.\ Temp.\ Phys.\  {\bf 184}, no. 1-2, 74 (2016).
  
  \bibitem{Edwards:2012}
  J.M.~Edwards, R.~O\rq{}Brient, A.T.~Lee, G.M.~Rebeiz, 
  IEEE Trans. Antennas Propag. , vol. 60, pp. 4082-4091, Sept. 2012
  
  \bibitem{Benson:2014qhw} 
  B.~A.~Benson {\it et al.} [SPT-3G Collaboration],
  Proc.\ SPIE Int.\ Soc.\ Opt.\ Eng.\  {\bf 9153}, 91531P (2014)
  
  \bibitem{Kermish:2012eh} 
  Z.~Kermish {\it et al.},
  Proc.\ SPIE Int.\ Soc.\ Opt.\ Eng.\  {\bf 8452}, 1C (2012)
  
\bibitem{Suzuki:2015zzg} 
  A.~Suzuki {\it et al.} [POLARBEAR Collaboration],
  J.\ Low.\ Temp.\ Phys.\  {\bf 184}, no. 3-4, 805 (2016)
  
  \bibitem{Matsumura:2016sri} 
  T.~Matsumura {\it et al.},
  J.\ Low.\ Temp.\ Phys.\  {\bf 184}, no. 3-4, 824 (2016).
  
  
  
  

\end{thebibliography}
\end{document}